# Emerging negative Poisson's ratio driven by strong intralayer interaction response in rectangular transition metal chalcogenides

Linfeng Yu[1], Yancong Wang[1], Xiong Zheng[1], Huimin Wang[2], Zhenzhen Qin[3], and Guangzhao Qin[1],*

[1] *State Key Laboratory of Advanced Design and Manufacturing for Vehicle Body, College of Mechanical and Vehicle Engineering, Hunan University, Changsha 410082, P. R. China*

[2] *Hunan Key Laboratory for Micro-Nano Energy Materials & Device and School of Physics and Optoelectronics, Xiangtan University, Xiangtan 411105, Hunan, China*

[3] *School of Physics and Microelectronics, Zhengzhou University, Zhengzhou 450001, China*

**Abstract**

Auxetic behavior quantified by the negative Poisson's ratio (NPR) is commonly attributed to geometry evolution with re-entrant mechanism or other mechanical factors, which is thought to be independent of electronic structures. Thus, searching for electronic effect dominated auxetic behavior is challenging. Herein, from *state-of-the-art* first-principles calculations, by studying a class of two-dimensional (2D) transition metal chalcogenides (TMCs), namely $X_2Y_2$-type (X=Cu, Ag, Au; Y=O, S, Se) rectangular TMCs (R-TMCs), we identify that the monolayer R-$Cu_2Se_2$ unconventionally demonstrates a structure-independent anisotropic NPR. In contrast, the NPR is absent in other R-TMCs. The emerging NPR is attributed to the strong strain response of intralayer interaction in R-$Cu_2Se_2$, which can be traced to the lone pair electrons and weak electronegativity of Se atoms under multi-orbital hybridization. The emerging NPR would make R-$Cu_2Se_2$ a promising candidate in electronics protection, and our study would provide valuable clues and useful guidance for designing advanced auxetic materials.

* Author to whom all correspondence should be addressed. E-Mail: gzqin@hnu.edu.cn

# 1. Introduction

The last years have witnessed the prosperity of auxetic materials. As a functional mechanical phenomenon, auxetic effect means that a material expands (contracts) when it is stretched (compressed) *i.e.* negative Poisson's ratio (NPR). Auxetic materials have excellent mechanical properties such as enhanced toughness, shear resistance and vibration absorption[1]. Hence, they has greatly potential applications in medicine, defense, tougher composites and other broad fields[2–4]. In classical elastic theory, Poisson's ratio ranges from -0.5 to 1[5], and then the NPR is theoretically reasonable in three-dimensional (3D) continuous materials like bulk auxetic structures[6,7] and metals[8,9]. With the explosion of two-dimensional (2D) materials, the excellent properties of 2D materials have been driving its rapid development. Outstanding representatives include graphene-like materials[10–13], transition metal chalcogenides (TMCs) [14–16], and transition metal halide[17–19] *etc*. it is expected to find this novel NPR phenomenon in 2D materials.

The intrinsic NPR in 2D materials refers to the auxetic effect that manifests without external influence, which can be divided into five behaviors[20] as follows: (1) *In-plane* NPR behavior; (2) *Out-of-plane* NPR behavior; (3) Bidirectional NPR behavior; (4) *In-plane* half-NPR behavior[21]; (5) *Out-of-plane* half-NPR behavior. However, most 2D materials tend to have a positive Poisson's ratio (PPR), and their NPR must be activated by external techniques. For instance, the NPR in 2D graphene-like compound[22,23,24] can be activated through strain engineering. In addition, oxidation[25], vacancy defects[26], and cutting into nanoribbons[27] are also reported to activate NPR. In recent years, many efforts have been devoted to searching for intrinsic NPR in 2D materials based on specific lattice structure, including puckered -materials[28,29], square-materials[15,30], and other configurations[21,31–33]. However, the origin of intrinsic NPRs is usually attributed to the unique geometric structures with a re-entrant mechanism or other mechanical factor, regardless of its electronic properties. For instance, the NPR of puckered materials such as black phosphorene originates from the re-entrant mechanism[28,29], and the recently reported NPR of wurtzite monolayers originates from the folded in-plane triangular stacking[34]. Distinct from above auxetic materials, an unusual NPR behavior dominated by electronic effects is found in 1T-TMCs[35] by *Yu et al* inadvertently in the form of filled electron orbitals. Combined with other remarkable properties of such NPR materials could lead to unexpected multi-functionalities. Since then, there is no similar NPR materials dominated by electronic effects has been found in any other configuration and the lack of cases provides poor insight into this unique behavior, which arise a thought-provoking question whether there are other special NPR mechanisms beyond 1T-TMCs or any auxetic configurations without re-entrant.

In this study, we report a novel 2D TMCs with intrinsic NPR behavior based on first-principles methods. The difference from other reported auxetic materials lies in their mechanical structure and more in the microscopic origin of their Poisson's ratio behavior. All TMCs monolayers are of rectangular $X_2Y_2$ type (X=Cu, Ag, Au and Y=O, S, Se), named R-$X_2Y_2$. Most importantly, a structure-independent NPR behavior is found in monolayer R-Cu$_2$Se$_2$. Unlike previously reported NPRs dominated by re-entrant mechanisms, such behavior cannot gain insight from geometric behavior because while other R-X$_2$Y$_2$ configurations are not found. This emerging NPR in R-X$_2$Y$_2$ monolayers can be explained by their distinct response strength of intralayer interactions. In addition, the stability of R-X$_2$Y$_2$ monolayers is confirmed to be dynamically, thermally and mechanically stable and they exhibit promising electrically semiconducting and optical absorption properties. Auxetic properties combined with these promising optical and electrical properties. properties make R-TMCs a promising candidate for protecting electronics.

## 2. Results and discussion

### 2.1. Structure of R-X$_2$Y$_2$ and the stability verification

As shown in Fig. 1(a), the primitive cell of R-X$_2$Y$_2$ monolayers contains two X (X=Cu, Ag and Au) atoms and two Y (Y=O, S and Se) atoms, which is different from XY$_2$-type 1T (or 2H)-TMCs. From the side view, R-X$_2$Y$_2$ monolayers have a zigzag-shaped washboard configuration, and they do not exhibit mirror symmetry (like 2H-TMCs) but centrosymmetric (like 1T-TMCs). Interestingly, R- X$_2$Y$_2$ monolayers inherit the non-metal shielding properties in traditional TMCs (1T and 2H phase), namely, the metal X atoms are in the middle layer and the non-metal Y atoms are in the outer layer. In addition, the X atom in the middle of the layer forms a two- and four-coordinate bonding configuration with the Y atom, respectively, while the two Y atoms only form an equivalent three-coordinate configuration. Fig. 1(b) further demonstrates the localized symmetric atomic schematic involving NPR evolution. The labels *a*, *b*, *h* and *θ* represent the lattice constant along the *x* and *y* directions, the buckling height and characteristic angle, respectively, which are listed in Table 1. The unique buckling structure reduces atomic Y-Y repulsion, and thus R-X$_2$Y$_2$ may be relatively stable.

The stability of R-X$_2$Y$_2$ monolayers can be confirmed based on phonon dispersion as shown in the supplementary Fig. S1, which reflects the dynamic stability of the system. Considering that the NPR only appears in monolayer R-Cu$_2$Se$_2$, we only focus on the results of monolayer R-Cu$_2$Se$_2$. The phonon dispersion of monolayer R-Cu$_2$Se$_2$ is plotted in Fig. 1(c), where no imaginary frequencies appear in the acoustic phonon branch. Thus, monolayer R-Cu$_2$Se$_2$ is dynamically stable. Furthermore, *ab initio molecular dynamics* (AIMD)[36,37] simulations are performed to assess thermal stability, which are shown in the supplementary Fig. S2. The stable temperature distribution at 300 K and a snapshot of the configuration after the AIMD simulation indicate its stability in room temperature. Mechanically, the stability of monolayer R-Cu$_2$Se$_2$ need to meet the Born−Huang criteria[38], *i.e.*, $C_{11}$, $C_{12}$, $C_{66}$ > 0 and $C_{11}+C_{22}$ > $2C_{12}$, where $C_{ij}$ represents elastic constant. The calculated values of $C_{11}$, $C_{22}$, $C_{12}$, and $C_{66}$ of R-X$_2$Y$_2$ monolayers are listed in Table 1, which satisfy mechanical stability. Noted that the $C_{11}$, $C_{22}$, $C_{12}$, and $C_{66}$ of 38.27, 2.88, -0.71 and 0.84 N/m for monolayer R-Cu$_2$Se$_2$ satisfies the criteria. These results strongly demonstrate the stability of monolayer R-Cu$_2$Se$_2$. More mechanical information is found in Supplementary NoteS1, Table S1 and Fig S4.

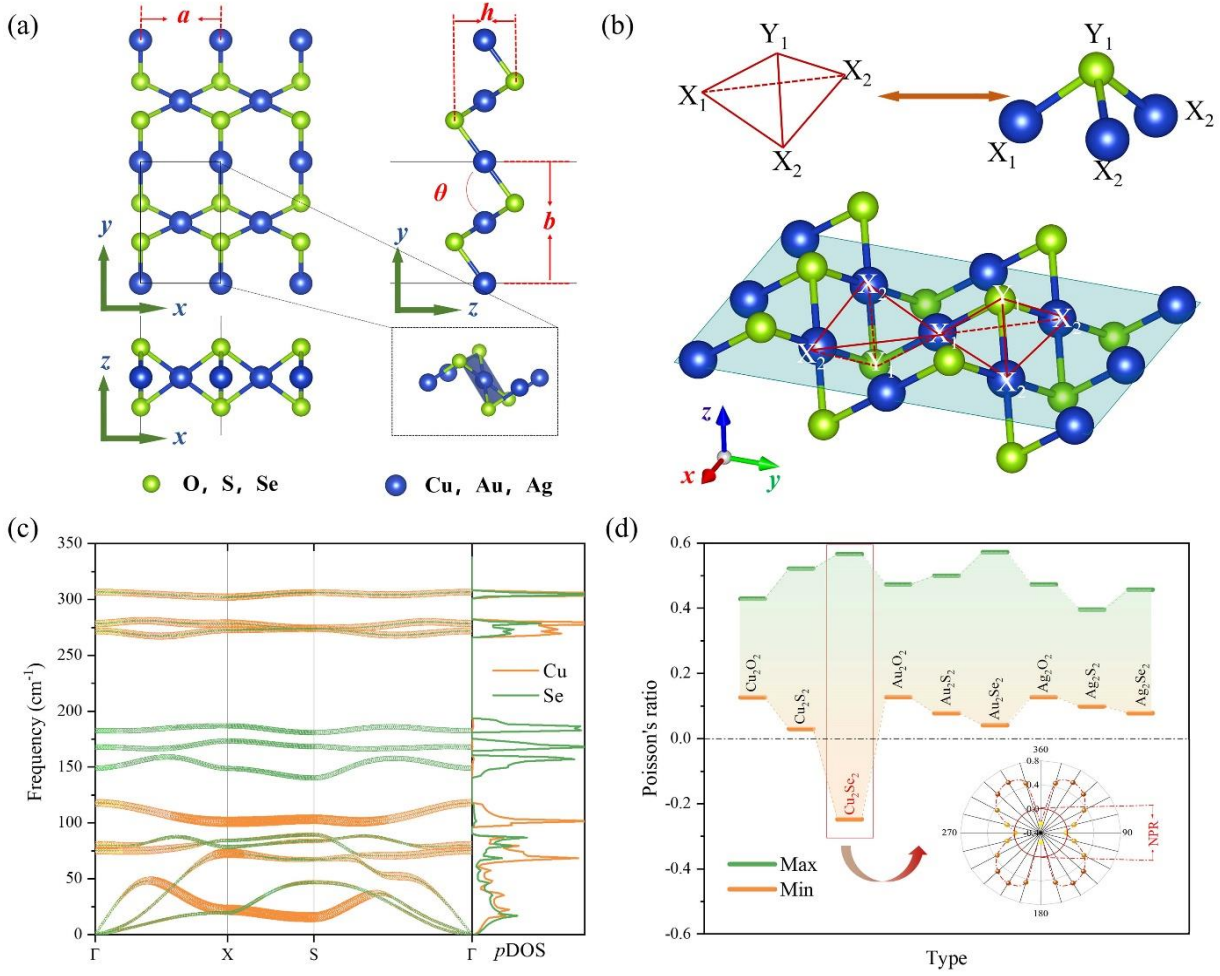

Figure 1. (a) The views of the geometry structures of R- X$_2$Y$_2$ (X=Cu, Ag, Au and Y=O, S, Se). The gray dashed line marks the crystal lattice. (b) Local lattice morphology and atoms with symmetrical motion in R-X$_2$Y$_2$. (c) Phonon dispersion and partial density of states (pDOS) of monolayer R-Cu$_2$Se$_2$ with spitted contribution from different atoms. (d) The highest and lowest Poisson's ratio of R-X$_2$Y$_2$ monolayers. The inset shows the direction dependence of Poisson's ratio for monolayer R-Cu$_2$Se$_2$.

Table 1. Comparison of basic physical properties for R-X$_2$Y$_2$ (X=Cu, Ag, Au; Y=O, S, Se) monolayers. Lattice constant *a*, *b*; Buckling height *h*; Characteristic angle *θ*; Band gap from HSE06 ($E_{HSE06}$) and PBE ($E_{PBE}$) functionals; Elastic constant C$_{ij}$ (*i, j*=1, 2, 6).

|  | a (Å) | b (Å) | h (Å) | θ (°) | $E_{HSE06}$ (eV) | $E_{PBE}$ (eV) | C$_{11}$ (N/m) | C$_{12}$ (N/m) | C$_{22}$ (N/m) | C$_{66}$ (N/m) |
|---|---|---|---|---|---|---|---|---|---|---|
| Cu$_2$O$_2$ | 2.80 | 5.66 | 2.02 | 136.14 | 0.79 | _ | 68.45 | 8.65 | 20.92 | 8.88 |
| Cu$_2$S$_2$ | 3.35 | 5.50 | 2.29 | 97.20 | 1.27 | 0.64 | 40.74 | 1.23 | 7.55 | 2.37 |
| Cu$_2$Se$_2$ | 3.54 | 5.41 | 2.65 | 87.34 | 1.11 | 0.6 | 38.27 | -0.71 | 2.88 | 0.84 |
| Ag$_2$O$_2$ | 3.05 | 6.28 | 1.22 | 135.36 | 1.25 | 0.17 | 52.12 | 6.65 | 17.70 | 6.20 |
| Ag$_2$S$_2$ | 3.57 | 6.07 | 2.45 | 98.57 | 1.60 | 0.98 | 27.37 | 2.14 | 7.39 | 2.40 |
| Ag$_2$Se$_2$ | 3.76 | 6.05 | 2.76 | 91.03 | 1.27 | 0.86 | 22.70 | 0.94 | 4.73 | 1.28 |
| Au$_2$O$_2$ | 3.10 | 6.06 | 1.37 | 128.86 | 1.02 | 0.18 | 61.66 | 10.56 | 25.96 | 10.64 |
| Au$_2$S$_2$ | 3.54 | 6.26 | 2.26 | 105.6 | 2.19 | 1.35 | 41.26 | 4.10 | 13.05 | 5.40 |
| Au$_2$Se$_2$ | 3.72 | 6.34 | 2.55 | 99.26 | 1.79 | 1.21 | 34.56 | 2.70 | 9.23 | 3.37 |

## 2.2. Emerging NPR in R-Cu$_2$Se$_2$

Based on mechanical stability, mechanical properties of the R-X$_2$Y$_2$ monolayers are investigated. In-plane Poisson's ratio is obtained as follows[31]:

$$v(\theta) = \frac{C_{12}sin^4\theta - Bsin^2\theta cos^2\theta + C_{12}cos^4\theta}{C_{11}sin^4\theta - Asin^2\theta cos^2\theta + C_{22}cos^4\theta} \quad (1)$$

where $A = (C_{12}C_{22} - C_{12}^2)/C_{66} - 2C_{12}$ and $B = C_{11} + C_{22} - (C_{12}C_{22} - C_{12}^2)/C_{66}$. Fig. 1(d) shows the maximum and minimum values of Poisson's ratio for R-TMCs. Compared with other R-X$_2$Y$_2$ binary compounds, R-Cu$_2$Se$_2$ has the widest Poisson's ratio range, which is between -0.3 and 0.6. Most interestingly, an emerging NPR is found in R-Cu$_2$Se$_2$ despite having a similar structure to other binary R-X$_2$Y$_2$. The embedding of Fig. 1(d) is the Poisson's ratio of R-Cu$_2$Se$_2$ as a function of angle $\theta$. The Poisson's ratio is the smallest and negative along the 0 and 180 degree directions, while the largest positive Poisson's ratio occurs at ~30, ~150, ~210, and ~330 degrees. Monolayer R-Cu$_2$Se$_2$ exhibits a strong anisotropic NPR along the orthogonal direction. The high NPR of -0.27 comes from the *y* direction, which is higher than α-phosphorene (-0.027)[28], tetra-silicene (-0.055)[39], WN$_4$ (0.113)[40], penta-graphene (0.068)[41] and Be$_5$C$_2$ (-0.16)[42]. In addition, the orthogonal anisotropic in-plane Poisson's ratio is rarely found in other anisotropic configurations, such as hinge-structures materials[28,43], and borophane[32]. Most importantly, this emerging NPR is independent of traditional re-entry mechanisms, since NPR behavior is not found in other R-X$_2$Y$_2$ monolayers. Such an effect cannot be intuitively explained geometrically and a similar phenomenon is reported in 1T-TMCs[35] due to the strong coupling between specific orbitals. The difference is that R-X$_2$Y$_2$ of the same main group has similar orbital hybridization, but only R-Cu$_2$Se$_2$ exhibits NPR behavior. The emerging NPR in R-Cu$_2$Se$_2$ is expected to provide new valuable clues for the regulation of electronic effects on NPR behavior.

## 2.3. Anisotropy of the NPR

Furthermore, orthogonal forced strains are applied in the *x* and *y* directions to reveal the anisotropic NPR behavior for R-Cu$_2$Se$_2$, respectively. The response of monolayer R-Cu$_2$Se$_2$'s lattice parameters to uniaxial strain is plotted in Fig. 2. The Poisson's ratio is defined as[44]

$$v_1 = -\partial\varepsilon_1/\partial\varepsilon_2, \tag{2}$$

where $\varepsilon_1$ is the response strain driven by forced strain $\varepsilon_2$. The intrinsic Poisson's ratio $v_1$ can be obtained by fitting $\varepsilon_1 = -v_1\varepsilon_2 + v_2\varepsilon_2^2 + v_3$. The behavior for a larger strain range (-6%-6%) is provided in Supplementary Fig. S3. Here, as shown in Fig. 2(a) and (b), the response lattice constant increases linearly with forced strain, indicating that the lattice expands when it is stretched, *i.e.*, NPR. The intrinsic Poisson's ratios in the *y* direction and *x* direction are -0.25 and -0.07, respectively, which are in good agreement with the results based on the elastic constant calculation. The primitive cell of monolayer R-Cu$_2$Se$_2$ contains two Cu atoms and two Se atoms, but the displacement mode of the two Se atoms is equivalent due to symmetry. Thus, the mechanical explanation of NPR can be understood by tracing the evolution of the tetrahedron composed of Cu$_1$, Cu$_2$ and Se$_1$ atoms as shown in Fig. 1(b) and Fig. 2(c, d). As shown in Fig. 2(c), a uniaxial forced strain will increase the distance between its adjacent Cu$_1$ and Cu$_2$ atoms, inducing the Se$_1$ atom to move downward when it is applied along the *x* direction. Critically, angle $\theta$ increases, contributing to NPR. Similarly, strain along the *y* direction induces the downward movement of Se$_1$ atoms as shown in Fig. 2(d), which causes the Cu$_1$ atoms to move to both sides of the *x*-axis, and eventually produces an NPR. The origin of anisotropic NPR lies geometrically in the flattening of the tetrahedron due to uniaxial strain.

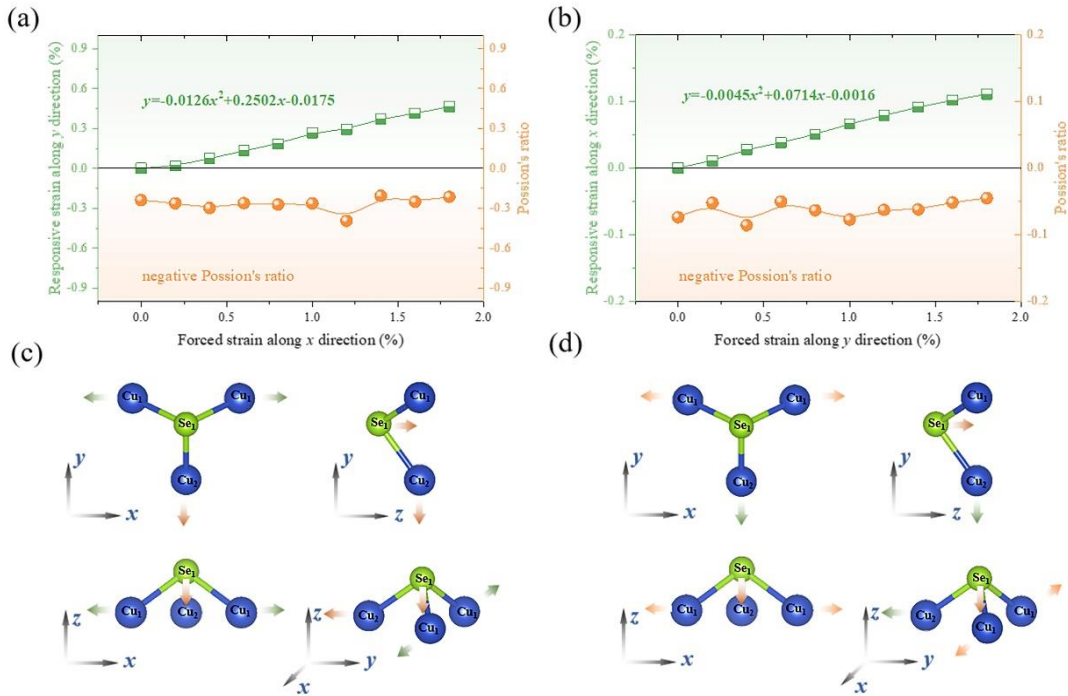

Figure 2. Mechanical response driven by strains along the (a) *x* and (b) *y* directions, respectively. The localized atomic displacements corresponding to the negative Poisson's ratio along the (c) *x* and (d) *y* directions, respectively. The green and red arrows indicate forced and responsive strain-driven atomic displacements, respectively.

## 2.4. The mechanism underlying NPR

Noticed, the geometric evolution is not sufficient to explain the emerging NPR, as only R-$Cu_2Se_2$ appears NPR but not in other R-$X_2Y_2$. To further explain such emerging NPR, we take the geometrical and electronic responses of monolayer R-$Cu_2Se_2$ and R-$Ag_2Se_2$ when strain is applied along the *x*-direction as an example. The geometric response along the *y*-axis is related to the projected length *l* of R-$X_2Y_2$ on the side, which further depends on the geometrical change of the lattice atoms on the projected triangle. Hence the key parameters are obtained, namely the feature lengths $l_1$ and $l_2$ and the feature angle *θ*. When tensile strain is applied, two competing geometric modes are extracted as shown in Fig. 3 (e): (1) geometrical length (GL) mode, where only the bond length shortens; (2) geometrical angle (GA) mode, meaning only the angle changes. The contributions of the two modes to the Poisson ratio are opposite and competing, *i.e.* the GA mode tends to have a NPR while the GL mode contributes a positive Poisson's ratio (PPR). Interestingly, the geometric responses of monolayer R-$Cu_2Se_2$ and R-$Ag_2Se_2$ exhibit a consistent behavior as shown in Fig. 3 (a) and (b), namely feature angle *θ* increases and feature lengths $l_1$ and $l_2$ decrease. Similar geometric evolution stems from their same geometric configuration and the same main group of elements. In geometric evolution, the emerging NPR originates from the GA mode (the increase of feature angle *θ*). The difference is that the angle *θ* increases faster with strain in monolayer R-$Cu_2Se_2$ compared to that in monolayer R-$Ag_2Se_2$, which overcomes the GL mode that tends to produce a PPR. Compared to other R-$X_2Y_2$, monolayer R-$Cu_2Se_2$ has relatively strong strain response of intralayer interactions due to the geometrically smallest characteristic angle *θ* as shown in Fig. 3(c). A smaller characteristic angle means it is more likely to produce an increase in response. This strong intralayer interaction response originates from the overlapping of wave functions within the layers, which can be visually reflected in the electron localization function (ELF) as shown in Fig. 3(f–i). When tensile strain is applied, the overlap of electron clouds between Se atoms in monolayer R-$Cu_2Se_2$ [Fig. 3(f) and (g)] is deepened while the overlap in R-$Ag_2Se_2$ is relatively weak [Fig. 3(h) and (i)], which leads to stronger lateral repulsion within the layer.

Thereby a stronger intralayer interaction response is found in R-Gu$_2$Se$_2$, which is reflected in the angle $\theta$ increasing faster (stronger GA mode). To quantify the strength of intralayer interaction response, the integrated Crystal Orbital Hamilton Population (ICOHP) between the X$_1$ and X$_2$ atoms is extracted as shown in Fig. 3(d). When the strain increases by 4%, the ICOHP of monolayer R-Cu$_2$Se$_2$ increases by more than 7%, which is significantly faster than that of monolayer R-Ag$_2$Se$_2$ (no more than 2%). Faster growth in the ICOHP means a stronger intralayer interaction response. Therefore, the strong intralayer response drives the faster increase of the characteristic angle $\theta$ in monolayer R-Cu$_2$Se$_2$, ultimately leading to structure-independent NPR behavior.

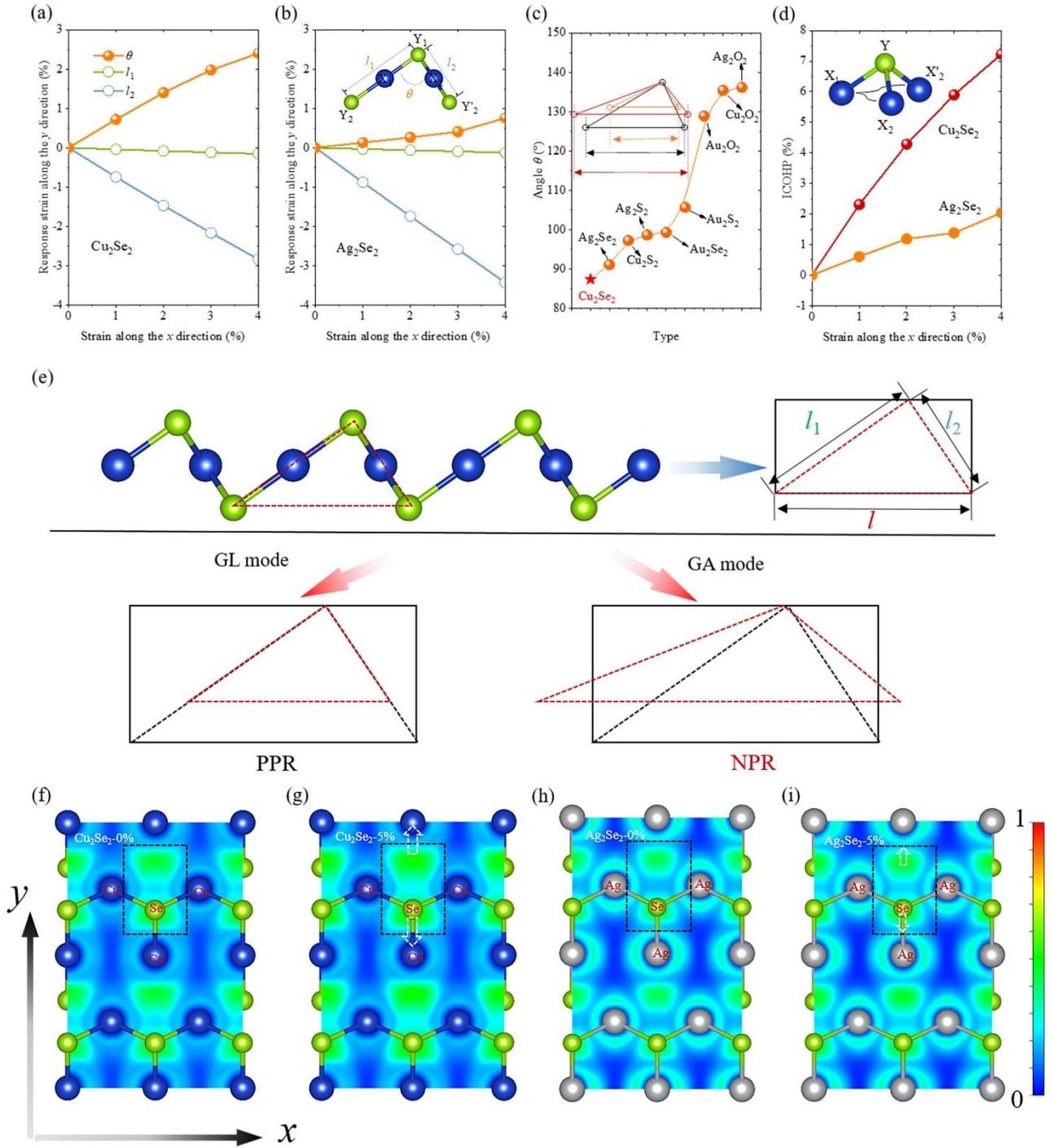

Figure 3. The origin of the NPR. Geometric response driven by strains along the *x* direction in (a) R-Cu$_2$Se$_2$ and (b) R-Ag$_2$Se$_2$. (c) Comparison of characteristic angles in R-X$_2$Y$_2$ with different components. (d) Contrast of changes in ICOHP with strain for R-Cu$_2$Se$_2$ and R-Ag$_2$Se$_2$. (e) Two competing modes in the evolution of NPR, including geometrical length (GL) and geometrical angle (GA) modes. (f, g) The electron localization function (ELF) of Cu$_2$Se$_2$ when the strain is (f) 0% and (g) 5%. (h, i) The ELF of Ag$_2$Se$_2$ when the strain is (h) 0% and (i) 5%.

## 2.5. Fundamental understanding of NPR from orbital evolution

To gain further insight into the emerging NPR in this novel $X_2Y_2$-type R-TMCs monolayers, fundamental electronic properties of the R-$X_2Y_2$ monolayers need to be investigated. Further, the HSE06 functional is considered to obtain accurate band gap, which are higher than the results of the PBE functional as shown in Fig. 4(a) and Table. 1. Differently, all R-$X_2Y_2$ monolayers exhibit semiconducting characteristics based on the HSE06 functional. Among them, monolayer R-$Au_2S_2$ exhibits the largest band gap of 2.19 eV while the smallest band gap of monolayer R-$Cu_2O_2$. The HSE06-based band gap of the R-$X_2Y_2$ monolayers is in the range of 0.7 to 2.2 eV, which fits well within the ideal band gap (0.9-1.6 eV) for optical absorption as shown in the iridescent area in Fig. 4(a). Based on the HSE06 functional, monolayer R-$Cu_2Se_2$ exhibits an indirect band gap of 1.16 eV as a semiconductor. As shown in Fig. S6, R-$X_2Y_2$ monolayers have strong absorption peaks in the visible range of 1.62 eV~3.11 eV, which means that R-$X_2Y_2$ monolayers have outstanding phonon absorption ability. Besides, R-$X_2Y_2$ monolayers also have strong absorption peaks in the ultraviolet region (>3.11 eV), indicating their great potential in optical devices. More information can be provided in the supplementary Note S2 and Fig. S6. Combined with this emerging NPR, the promised electrical and optical properties enable R-$X_2Y_2$ to meet functional requirements for practical applications.

Further, the orbital evolution of R-$Cu_2Se_2$ is plotted in Fig. 4(b) to demonstrate fundamental understanding for emerging NPR. The valence band maximum (VBM) is located between the high symmetry point path S-Y, and the conduction band minimum (CBM) is located between Γ-X. Fig. 4(c) shows the 3D images of VBM and CBM. The conduction band with the lowest energy forms two valley peaks ($C_1$ and $C_2$) with a little energy difference of 0.19 eV, which are located in the Γ-X and S-Y paths respectively. Meanwhile, the VBM along S-Y exhibits a quasi-flat energy band, forming a quasi-direct band gap of 1.35 eV. The quasi-flat band leads to a sharp peak in partial density of states (*p*DOS). The VBM are mainly contributed by Se-*p*, Cu-*p* and Cu-*d* orbitals while the CBM is dominated by Se-*p* and Cu-*d* orbitals. In the entire bonding region, the contribution of Se-*p* and Cu-*d* orbitals is much higher than that of other orbitals. Noted that the energy of the *s* electron in the Se atom is much lower than the bonding electrons than Cu atom, forming a lone pair of electrons as shown in the inset in the lower right corner of Fig. 4(b). The weak *sp*$^3$ hybridization is captured and further hybridized with the *d* orbital to produce a more complex multi-orbital effect. Hence, the Se atoms in the pyramidal coordination center form complex multi-orbital hybridizations during NPR evolution, which are more likely to promote smaller bond angles than simple *sp* or *sp*$^2$ hybridizations. In addition, the electron cloud of the lone pair *s* electrons in the central Se atom is hypertrophic, which has a large repulsive force for the bonding electron pair. Thus, the lone pair *s* electrons can make the bonding electron pair closer to each other, and the bond angle is compressed and becomes smaller. Furthermore, the central Se atom has a smaller electronegativity of 2.55 compared to 2.58 and 3.44 for the O and S atoms. Hence, the bonding electron is farther away from the Se atom, the repulsion between the bonding electron pair becomes smaller, and the bond angle becomes smaller. The lone pair *s* electrons and weak electronegativity in Se atoms can be intuitively revealed through the petal-shaped electron cloud in the ELF shown in Fig. 3 (d) and (e). The strong lone-pair electrons and weak electronegativity of Se atom leads to a smaller characteristic angle of R-$Cu_2Se_2$ [Fig. 3(c)]. Thus, when subjected to tensile strain, the GA mode is easier to respond [Fig. 3(a)] and the stronger intralayer interaction response [Fig. 3(d)] can be captured, ultimately contributing to the unique NPR behavior.

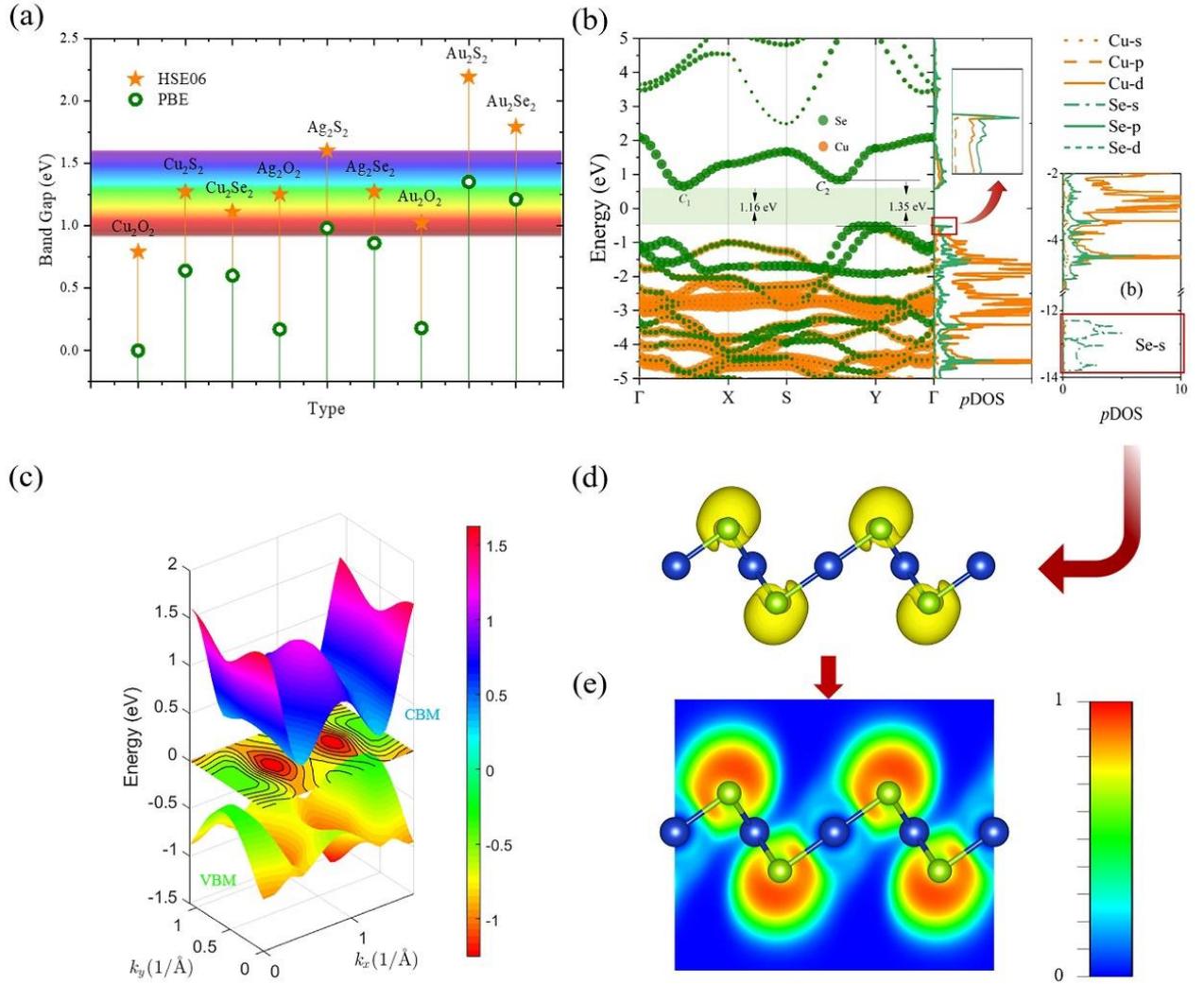

Figure 4. Electronic properties of R-$X_2Y_2$ (X=Cu, Ag, Au; Y=O, S, Se) monolayers. (a) The band gaps of the R-$X_2Y_2$ monolayers from the HSE06 and PBE functionals. The rainbow band means the ideal optical absorption band gap. (b) The energy band structure of monolayer R-$Ce_2Se_2$. (c) The 3D CBM and VBM of monolayer R-$Ce_2Se_2$. (d) 3D electron localization function (ELF) for monolayer R-$Ce_2Se_2$ (isosurface of 0.8). (e) 2D electron localization function of monolayer R-$Ce_2Se_2$.

## 3. Conclusion

In summary, we have proposed a novel 2D TMCs (R-TMCs) with NPR and performed a comprehensive investigation of the Poisson's ratio behavior in $X_2Y_2$-type (X=Cu, Ag, Au and Y=O, S, Se) rectangular TMCs monolayers. When strain is applied, only monolayer R-$Cu_2Se_2$ has an intrinsic NPR behavior. Unlike the previously reported re-entry mechanism-dominated NPR behavior, this behavior can be attributed to a strong strain response of intralayer interaction driven by electronic effects. Insights into the origin of NPR can be obtained to trace key parameters (length $l_1$, $l_2$ and angle $\theta$) in geometric evolution, which can be further explained by the response of two competing modes: GA mode (dominated by angle $\theta$) and GL mode (dominated by length $l_1$, $l_2$). All R-$X_2Y_2$ monolayers exhibit the same evolutionary path, but the response strength of R-$Cu_2Se_2$ is stronger due to the smaller angle $\theta$. The strong intralayer interaction response originates from the lateral repulsive forces generated by the accumulated intralayer charges, and is further quantified by ELF and ICOHP, which can be further traced to the lone pair electrons and weak electronegativity of Se atoms under multi-orbital hybridization. The understanding of this unique NPR mechanism can provide valuable clues and guidance for the discovery or design of novel auxetic materials. To our knowledge, such NPR behavior has not been reported in other copper selenide phases, although there is ubiquitous copper compounds in nature's abundant mineral resources or even copper selenides in different dimensions have been synthesized experimentally. Combining the promising electrical and optical properties, the unique auxetic properties enable monolayer R-$Cu_2Se_2$ to lead to multi-functionality.

## 4. Computational Details

All first-principles calculations are performed based on density functional theory (DFT) utilizing the Vienna *ab-initio* simulation package (VASP)[45]. Based on the Perdew–Burke–Ernzerhof (PBE)[46] functional, the kinetic energy cutoff of 740 eV and a Monkhorst-Pack[47] *q*-mesh of 21×14×1 are used for structural optimization until the energy accuracy of $10^{-6}$ eV and the Hellmann-Feynman force accuracy of $10^{-3}$ eV/Å. Phonon dispersion is calculated by constructing a 6×4×1 supercell with a 2×2×1 *q*-mesh based on the finite displacement difference method as implemented in *PHONOPY*[48]. The calculation of the elastic constant is based on the energy-strain method using *VASPKIT*[49]. The accurate energy band gap and optical properties are based on the hybrid HSE06 functional[50,51].

## ACKNOWLEDGEMENTS

This work is supported by the National Natural Science Foundation of China (Grant Nos. 52006057, 51906097, and 11904324), the Fundamental Research Funds for the Central Universities (Grant Nos. 531119200237 and 541109010001), and the State Key Laboratory of Advanced Design and Manufacturing for

Vehicle Body at Hunan University (Grant No. 52175013). The numerical calculations in this paper have been done on the supercomputing system of the National Supercomputing Center in Changsha.**AUTHOR CONTRIBUTIONS**

*G.Q.* supervised the project. *L.Y.* performed all the calculations and analysis. All the authors contributed to interpreting the results. The manuscript was written by *L.Y.* with contributions from all the authors.**References**